# Point defects in epitaxial silicene on Ag(111) surface

Hongsheng Liu[1], Haifeng Feng[2], Yi Du[2*], Jian Chen[3], Kehui Wu[3], Jijun Zhao[1,4*]

[1] *Key Laboratory of Materials Modification by Laser, Ion and Electron Beams (Dalian University of Technology), Ministry of Education, Dalian 116024, China*

[2] *Institute for Superconducting and Electronic Materials (ISEM), University of Wollongong, Wollongong, New South Wales 2525, Australia*

[3] *Beijing National Laboratory for Condensed Matter Physics and Institute of Physics, Chinese Academy of Sciences, Beijing 100190, China*

[4] *Beijing Computational Science Research Center, Beijing 100084, China*

**Abstract**

Silicene, a counterpart of graphene, has achieved rapid development due to its exotic electronic properties and excellent compatibility with the mature silicon-based semiconductor technology. Its low room-temperature mobility of ~100 cm$^2 \cdot$V$^{-1}$s$^{-1}$, however, inhibits device applications such as in field-effect transistors. Generally, defects and grain boundaries would act as scattering centers and thus reduce the carrier mobility. In this paper, the morphologies of various point defects in epitaxial silicene on Ag(111) surfaces have been systematically investigated using first-principles calculations combined with experimental scanning tunneling microscope (STM) observations. The STM signatures for various defects in epitaxial silicene on Ag(111) surface are identified. In particular, the formation energies of point defects in Ag(111)-supported silicene sheets show an interesting dependence on the superstructures, which, in turn, may have implications for controlling the defect density during the synthesis of silicene. Through estimating the concentrations of various point defects in different silicene superstructures, the mystery of the defective appearance of $\sqrt{13} \times \sqrt{13}$ silicene in experiments is revealed, and 4×4 silicene sheet is thought to be the most suitable structure for future device applications.

---

[*] Corresponding authors. Email: zhaojj@dlut.edu.cn (J. Zhao), ydu@uow.edu.au (Y. Du)

# 1. Introduction

Silicene, a monolayer of silicon atoms arranged in a honeycomb lattice, has attracted great attention in recent years.[1-5] In contrast to flat graphene, silicene possesses a low-buckled structure with a buckled height of about 0.44 Å.[6-7] Nevertheless, silicene exhibits excellent electronic properties[6-14] similar to those of graphene.[15-19] Its band structure exhibits a linear dispersion and shows characteristic massless Dirac fermions with the Fermi velocity of $10^5$–$10^6$ ms$^{-1}$.[6, 11, 14] Due to strong spin–orbit coupling (SOC), the quantum spin Hall effect may be observed in silicene in an experimentally accessible temperature regime.[12] A tunable band gap can be opened up to about 4 eV in silicene by applying a perpendicular electric field[20-21] and by hydrogenation[22-25], halogenation[26-27], and oxidation[28-29]. Owing to these excellent properties and easy integration into the current Si-based semiconductor technology, silicene holds great promise for future applications in nanoelectronic devices.

To date, silicene with various superstructures, including (4×4), ($\sqrt{13} \times \sqrt{13}$)R13.9°, ($\sqrt{7} \times \sqrt{7}$)R19.1°, ($2\sqrt{3} \times 2\sqrt{3}$)R30° with respect to Ag(111)[30-35], and ($\sqrt{3} \times \sqrt{3}$) with respect to silicene 1×1 lattice[10, 30, 36], have been fabricated on Ag(111) surfaces. Very recently, a silicene field-effect transistor (FET) was successfully fabricated following a growth-transfer-fabrication process, in which the silicene device was encapsulated by delamination with native electrodes.[37] Nevertheless, the measured carrier mobility at room temperature was only about 100 cm$^2$V$^{-1}$ s$^{-1}$, which is three orders of magnitude lower than that of perfect free-standing silicene (2.6 × 10$^5$ cm$^2$V$^{-1}$ s$^{-1}$ for electrons and 2.2 × 10$^5$ cm$^2$V$^{-1}$ s$^{-1}$ for holes)[11] and even lower than that of a monolayer MoS$_2$ FET (~200 cm$^2$V$^{-1}$S$^{-1}$)[38]. Such a low mobility may be attributed to structural defects in the silicene. Generally, defects are inevitable in two-dimensional (2D) materials and have a significant impact on their physical properties. Therefore, a deep understanding of defects is highly desirable before fabrication of large-scale high-quality silicene layers for device applications. Recently, several typical point defects, including Stone-Wales (SW) rotation, single and double vacancies (SVs and DVs), and silicon adatoms (Si-ad) in freestanding silicene have been systematically investigated using density functional theory (DFT) calculations, focusing on the geometries, energetics, and effects on electronic properties.[39-49] It was found that the SWs and DVs may induce small gaps in silicene, while the SV defect leads to a semimetallic-to-metallic transition in silicene. Using the molecular dynamics finite element method with Tersoff potential,

Le reported that a single defect would significantly reduce the fracture strength of a silicene sheet.[43] Moreover, vacancy defects can reduce the thermal conductivity and the thermal stability of silicene.[44-45]

Despite the above efforts, a comprehensive understanding of point defects in epitaxial silicene is still lacking. In particular, most previous calculations considered only freestanding silicene. Despite that many previous experiments have observed defective features in the atomic structures of the Ag(111)-supported silicene samples[33-34, 50-56], there is no direct experimental identification on the point defects in epitaxial silicene. To address this critical issue, here, we present a systematical exploration of various types of point defects, including SW, SV, DV, and adatom in epitaxial silicene on Ag(111) surface using scanning tunneling microscopy combined with first-principles calculations. The agreement between the simulated STM images and the measured ones clarifies the atomic structures of point defects in epitaxial silicene. The formation energies and possible diffusion behavior of defects in two common silicene superstructures are compared and their implications for the defect density in experimentally synthesized silicene sheets are discussed.

## 2. Results and discussion

Four kinds of point defects in epitaxial silicene, including SW, SV, DV, and Si-ad, were considered in this work. In addition, possible imperfections of the Ag(111) surface, including single vacancy (Ag-SV) and adatom (Ag-ad), were also explored. Since the 4×4 and $\sqrt{13} \times \sqrt{13}$ superstructures are the most commonly observed structures of epitaxial silicene on Ag(111) surfaces[30-35], we mainly focus on these two structures in this paper. In freestanding silicene, there are two sets of sub-lattices; thus, the configurations of point vacancies (SW, SV, DV) are not dependent on local positions in a silicene sheet. The buckling of Si atoms in a silicene sheet would be rearranged, however, when the silicene is deposited on an Ag(111) surface. Therefore, the structure of a point defect in epitaxial silicene on the Ag surface becomes position dependent. For example, different morphologies for a SV within one unit cell of a given superstructure can occur, depending on whether the missing Si atom comes from the buckled-up atoms or the buckled-down atoms. In view of this fact, several possible structures for each kind of point defect have been taken into account for different silicene superstructures. To characterize the thermodynamic stability of a point defect in silicene, we define its formation energy $E_{\text{form}}$ as:

$$E_{\text{form}} = E_{\text{defect}} + N \times \mu_{\text{Si}} - E_{\text{perfect}} \qquad (1),$$

where $E_{\text{defect}}$ and $E_{\text{perfect}}$ are the energies for defective and perfect silicene@Ag supercells, respectively. $N$ is the number of missing atoms in the defective silicene@Ag supercell. For Si-ad, $N$ is set to be −1. $\mu_{\text{Si}}$ is the energy of one silicon atom in its bulk phase.

**2.1 Structures of various defects in 4×4 silicene**

Atomic structures of some representative point defects in the 4×4 silicene superstructure are shown in Figure 1. The SW defect in a 4×4 silicene sheet, formed by a Si-Si bond rotation of 90°, has the same configuration as in freestanding silicene[39, 46, 48], which is composed of two pentagons and two heptagons (Figure 1a). The SW defect in different positions results in a similar structure regardless of small differences in the detailed buckled pattern and formation energy, as depicted in Figure S1(a-d) in the Supporting Information. It is noteworthy that the formation energy for SW in epitaxial silicene (1.354−1.544 eV) is lower than in previous predictions for freestanding silicene (2.09 eV[39], 1.64 eV[46], and 1.82 eV[48]), owing to the passivation effect of the Ag substrate[57].

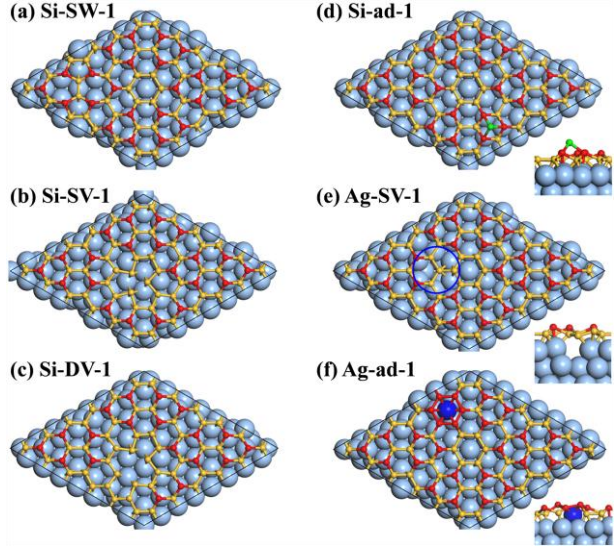

**Figure 1.** Atomic structures of several representative point defects in silicene and Ag substrate with a 4×4 silicene@Ag(111) superstructure. The small pictures in (d), (e), and (f) are the corresponding local side views around the defects. The sky-blue, dark-blue, yellow, red, and green balls represent Ag atoms, Ag adatoms, buckled-down Si atoms, buckled-up Si atoms, and Si

adatoms, respectively. The dark blue circle indicates the position of the Ag vacancy. The black rhombus indicates the simulation supercell.

For SV in a 4×4 silicene sheet, the configuration with three two-coordination silicon atoms (Si-SV-1, Figure 1b) is more energetically favorable ($E_{form}$ = 0.731 eV) than the reconstructed configuration (Si-SV-2, Figure S1e) proposed previously for freestanding silicene ($E_{form}$ = 1.059 eV)[39]. Again, the difference between freestanding and epitaxial silicene sheets can be ascribed to the passivation effect of the Ag surface. In freestanding silicene, the three dangling bonds created by one SV would significantly increase the energy. Hence, local structural reconstruction is needed to remove the dangling bonds. In epitaxial silicene, however, passivation by the Ag surface would stabilize the dangling bonds created by the defect, which is a common feature for silicene superstructures on metal substrates.[57-60] As a representative example, Si-DV-1 in 4×4 silicene is displayed in Figure 1c, which is similar to the DV-1 (5|8|5) proposed in our previous study.[39] Also note that SV and DV defects would severely influence the local buckling pattern of silicene, e.g. by reducing the height of the buckled-up silicon atoms by ~0.7 Å. For a Si adatom on 4×4 silicene, the preferred adsorption site is the hollow site of a hexagonal ring (Figure 1d), in which three silicon atoms are buckled up and the other three silicon atoms are buckled down, forming three Si-Si bonds with length of ~2.5 Å. A number of other metastable configurations for Si-SV, Si-DV, and Si-ad have been considered, and the details can be found in Supporting Information S1.

In this work, the effect on the morphology of epitaxial silicene due to possible imperfection of the Ag(111) surface, including single vacancy (Ag-SV) and adatom (Ag-ad) defects of Ag, was also taken into account. Combining DFT calculations and experimental STM observations, Satta and co-workers recently suggested that during the growth of silicene on Ag(111) surface, some Si atoms may penetrate the first Ag(111) layer and expel the Ag atoms.[61] Therefore, Ag-SV and Ag-ad may exist during the growth of silicene even though the prepared Ag(111) surface is flat enough. For Ag-SV-1 (Figure 1e), the missing Ag atom is right underneath the original buckled-up Si atom. After geometry relaxation, the original buckled-up Si atom moves down and becomes a buckled-down atom; while the other Si atoms are affected little by the Ag vacancy, with maximum displacement from their original locations by 0.05 Å. For Ag-ad, taking Ag-ad-1 (Figure 1f) as representative, one Ag adatom is located right underneath a hollow site of the silicene honeycomb lattice, lifting up the six silicon atoms in the hexagonal ring by about 1.33 Å. Several other types of Ag adatoms are displayed in Figure S3 of the Supporting Information.

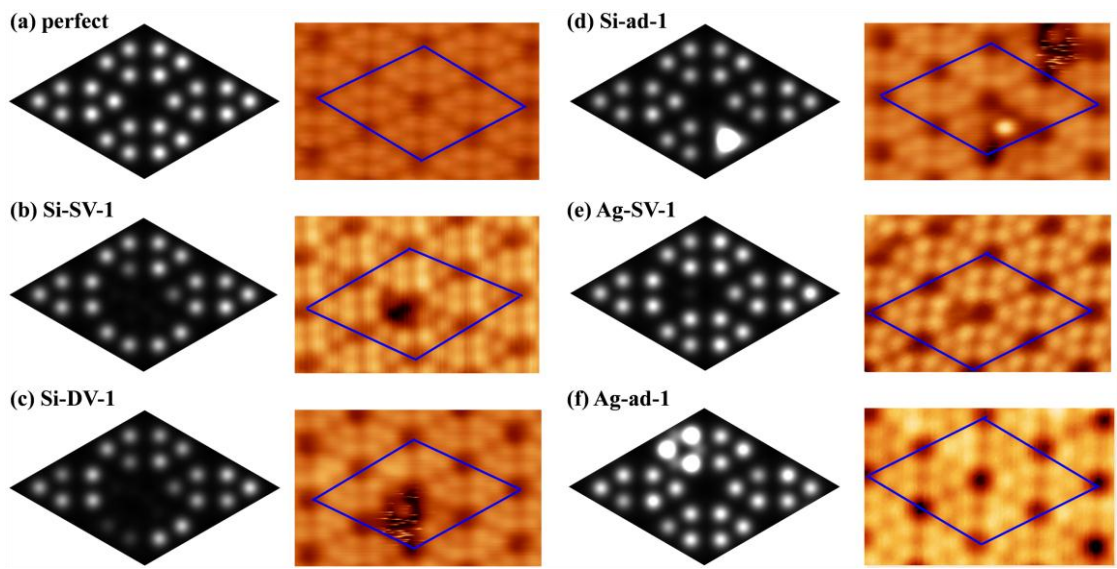

**Figure 2**. Simulated (left panels) and experimental (right panels) STM images for (a) perfect 4×4 silicene and for defective 4×4 silicene with (b) Si-SV-1, (c) Si-DV-1, (d) Si-ad-1, (e) Ag-SV-1, and (f) Ag-ad-1. The blue rhombus indicates the simulation supercell of 4×4 silicene. The bias voltages for the experimental STM images are 1.2 V for (a), 1.0 V for (b), 0.2 V for (c), 0.2 V for (d), −2.0 V for (e), and 0.8 V for (f), respectively.

Various structural defects in silicene sheets have been observed in previous experiments.[33-34, 50-56] The atomic structures of these defects are still unclear, however, due to the buckling structure of epitaxial silicene on metal substrates. To recognize the true defect structures in epitaxial growth of silicene and thus end the confusion, the STM images of all the defective silicene/Ag(111) superstructures constructed in this work were simulated and compared with the STM images from our own experiments. Firstly, the simulated STM image for perfect 4×4 silicene on Ag(111) agrees well with that obtained experimentally (Figure 2a). As shown in Figure 2b, the simulated STM image of Si-SV-1 exhibits a semilunar black hole, which coincides well with the experimental image. Careful observation tells us that three bright points are lost in the STM image of Si-SV-1 compared to the complete silicene lattice. The simulated STM image of Si-DV-1 (Figure 2c) shows a bigger black hole than that in Si-SV-1, since one more Si atom is missing. Si-ad-1, which is the most stable configuration for the Si adatom on 4×4 silicene, exhibits a large bright point in the STM image, as displayed in Figure 2d. Its apparent height is 1.45 Å higher than that of the buckled-up Si atoms (Figure 1d). The simulated STM image is also in accordance with the experimental image, in which one big bright point replaces the original three bright points in a triangle and the other bright points are nearly unaffected.

As discussed above, in 4×4 silicene with Ag-SV-1, one buckled-up Si atom in the original 4×4 silicene would become buckled down due to the missing Ag atom underneath. The local density of states will be affected, which results in the absence of an individual bright point in the simulated STM image for Ag-SV-1, as shown in Figure 2e. For Ag-ad-1, three original bright points arranged in a triangle are enhanced in the STM images, as shown in Figure 2f. This is because the Ag adatom lifts up the Si atoms above it and thus enhances the local density of states (LDOS). The excellent agreement between the simulated and experimental STM images for Ag-SV-1 and Ag-ad-1 confirms the presence of the structural defects in the Ag(111) surface underneath, in addition to those in the silicene sheet. This may also explain why it is hard to identify the point defects in epitaxial silicene grown on Ag(111) solely from the STM measurements in previous experiments.

## 2.2 Structures of various defects in $\sqrt{13} \times \sqrt{13}$ silicene

The atomic structures and simulated STM images of perfect $\sqrt{13} \times \sqrt{13}$ silicene are shown in Figure 3a. In our model of $\sqrt{13} \times \sqrt{13}$ silicene, there are four buckled-up Si atoms in one supercell, corresponding to four bright points in the STM image from both the DFT simulation and our experiment. Generally speaking, the SW, SV, and DV defects in a $\sqrt{13} \times \sqrt{13}$ silicene sheet share similar characteristics to those in a 4×4 silicene sheet, as displayed in Figure 3, and Figure S4 and Figure S5 in the Supporting Information. Si-SV-1 (Figure 3b), in which one buckled-up silicon atom is missing, is the most stable configuration for SV defects in $\sqrt{13} \times \sqrt{13}$ silicene, with ultralow formation energy of only 0.052 eV. On the other hand, the formation energies for Si-SV-2, Si-SV-3, and Si-SV-4 defects (Figure S5), in which the vacancy site corresponds to the buckled-down silicon atom, are much higher (0.148–0.534 eV). Compared to the buckled-down atom, the buckled-up silicon atom interacts more weakly with the Ag substrate and thus can be easily detached. On the other hand, the loss of this buckled-up silicon atom will help release strain in the epitaxial silicene sheet and, in turn, reduce the energy due to mismatch between the silicene lattice and the Ag substrate.

Similar to the case of the single vacancy, for a double vacancy with one missing buckled-up atom and one missing buckled-down atom, e.g., Si-DV-1 (Figure 3c), the formation energy is as low as 0.08 eV. Since there is only one lost buckled-up silicon atom at most, SV and DV defects in $\sqrt{13} \times \sqrt{13}$ silicene exhibit the same STM image, which is a perfect STM image with the lack of one bright point, as shown in Figure 3b and c. The excellent agreement between the simulated and experimental STM images for Si-SV and Si-DV clarifies that the missing bright point observed experimentally corresponds to the silicon vacancy defect. Note that the most preferred adsorption site for an Si adatom on $\sqrt{13} \times \sqrt{13}$ silicene is no longer the hollow site, but the top site, as displayed in Figure S4e and 4f of the Supporting Information.

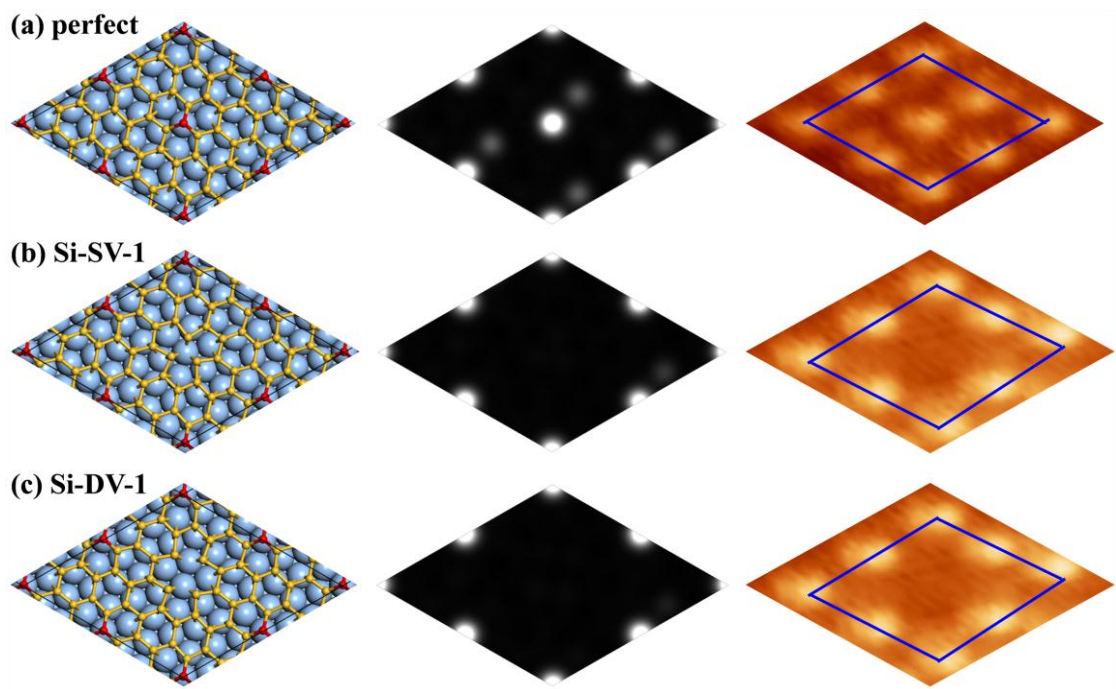

**Figure 3.** Atomic structures (left panels), and simulated (middle panels) and experimental (right panels) STM images of (a) perfect $\sqrt{13} \times \sqrt{13}$ silicene, (b) $\sqrt{13} \times \sqrt{13}$ silicene with silicon SV (Si-SV-1), and (c) $\sqrt{13} \times \sqrt{13}$ silicene with silicon DV (Si-DV-1). The sky-blue, yellow, and red balls represent Ag atoms, buckled-down Si atoms, and buckled-up Si atoms, respectively. The black rhombus in the atomic structures indicates the simulation supercell, which is also represented by the blue rhombus in the experimental STM images. The bias voltages for the experimental STM images are −0.5 V, and the tip current, $I_{tip}$ = 4 nA.

**2.3 Diffusion of defects**

At the growth temperature of silicene with 4×4 and $\sqrt{13}\times\sqrt{13}$ structures (typically 480–550 K), point defects may diffuse or aggregate, which would affect the distribution of defects and the final quality of the epitaxial silicene sheet. Considering this, we further explore the possible diffusion behavior of Si-SV and Si-ad in both 4×4 and $\sqrt{13}\times\sqrt{13}$ silicene using the climbing-image nudged elastic band (cNEB) method [62]. A schematic plot for the diffusion of a Si-SV in 4×4 silicene is shown in Figure 4. The initial configuration is set to be Si-SV-1, which is the most stable for an Si-SV, while the final configuration is Si-SV-2. The energy barrier for this diffusion is only 0.47 eV. Actually, when a Si-SV diffuses from Si-SV-1 to Si-SV-2, it can then diffuse from the Si-SV-2 to another Si-SV-1. In other words, a Si-SV can diffuse throughout the whole 4×4 silicene sheet with a maximal energy barrier of 0.47 eV. Also note that the diffusion barrier for an SV in freestanding silicene is only 0.12 eV, which is much lower than the present value due to the absence of the Ag substrate.[39]

The diffusion of an Si adatom in 4×4 silicene is plotted in Figure 5, in which the Si adatom migrates from a hollow site (Si-ad-1) to a top site (Si-ad-2) by overcoming an energy barrier of 0.41 eV and then diffuses to another top site, surmounting a barrier of 0.42 eV. Therefore, a Si adatom can diffuse from one of the most stable sites (hollow site) to another one by overcoming a maximal barrier of 0.42 eV, which is much lower than that in freestanding silicene (1.03 eV)[39].

To see how fast Si-SV and Si-ad can diffuse, we estimated the jump frequency, $p$, by[63]:

$$p \approx \nu \cdot \exp(-E_a/k_B T) \qquad (2),$$

where $\nu$ is a characteristic atomic vibrational frequency, $E_a$ is the activation energy for diffusion, $k_B$ is the Boltzmann constant, and $T$ is the temperature. The typical magnitude of the atomic vibration frequencies is about $10^{13}$ Hz. At 500 K, the jump frequency for Si-SV and Si-ad can be as high as $2.0 \times 10^8$ s$^{-1}$ and $6.2 \times 10^8$ s$^{-1}$, respectively. As a consequence, Si-SV and Si-ad can diffuse very fast in epitaxial 4×4 silicene sheet at the growth temperature.

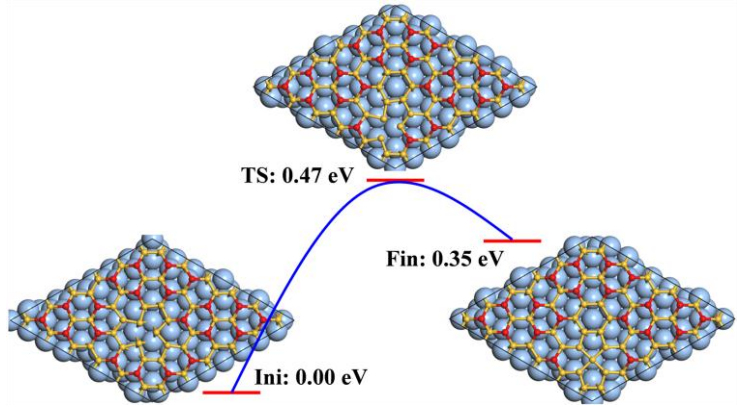

**Figure 4.** Schematic plots for diffusion of a SV defect in 4×4 silicene@Ag(111). The sky-blue, yellow, and red balls represent Ag atoms, buckled-down Si atoms, and buckled-up Si atoms, respectively. The black rhombus indicates the simulation supercell.

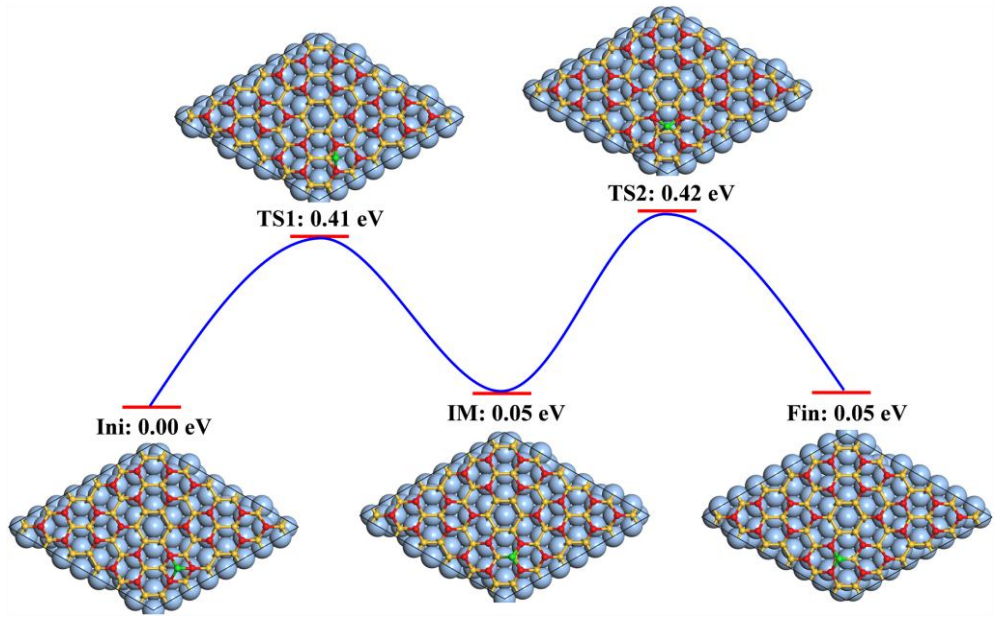

**Figure 5.** Schematic plots for diffusion of a Si-ad defect in 4×4 silicene@Ag(111). The sky-blue, yellow, red, and green balls represent Ag atoms, buckled down Si atoms, buckled up Si atoms, and Si adatoms respectively. The black rhombus indicates the simulation supercell.

For $\sqrt{13} \times \sqrt{13}$ silicene, the diffusion paths of Si-SV and Si-ad are plotted in Figure S6 and Figure S7 of the Supporting Information, showing the maximal activation energies of 0.76 eV and 1.04 eV, respectively. The corresponding jump frequencies at 500 K are estimated to be $2.0 \times 10^5$ s$^{-1}$ (Si-SV) and $3.8 \times 10^2$ s$^{-1}$ (Si-ad), respectively. Therefore, a Si single vacancy can still migrate in $\sqrt{13} \times \sqrt{13}$ silicene at an appreciable rate, but the diffusion is slower than in 4×4 silicene.

Meanwhile, diffusion of a Si adatom would be rather slow in $\sqrt{13} \times \sqrt{13}$ silicene.

Moreover, *ab initio* molecular dynamics (AIMD) simulations were performed for Si-SV in 4×4 and $\sqrt{13} \times \sqrt{13}$ silicene to examine the defect diffusion directly. Within the canonical *NVT* ensemble, the system temperature was set at 550 K, and the time step was 1 fs. The snapshot structures for Si-SV migration in 4×4 silicene are shown in Figure 6a. After 1.38 ps, the Si-SV moves to one side, with a silicon atom diffusing in the direction indicated by the red arrow. Then, this vacancy moves to a new position in the way indicated by red arrow at 2.12 ps. The Si-SV-2 configuration, which is the final state of the diffusion path in Figure 4, does not appear during the entire simulation time of 8 ps. This is because the Si-SV-2 configuration would occur at about 0.2 ns according to our estimated jump frequency of $5.2 \times 10^8$ s$^{-1}$ at 550 K, far beyond the time scale of the AIMD simulation. Nevertheless, we can still see the fast diffusion of Si-SV in the silicene sheet. For $\sqrt{13} \times \sqrt{13}$ silicene, the diffusion of a Si-SV is more evident, as shown in Figure 6b. At 2.71 ps, the Si-SV has already migrated far away from its original position. Then, the configuration of Si-SV at 2.71 ps is just the same as for Si-SV-3, which is shown in Figure S5b of the Supporting Information.

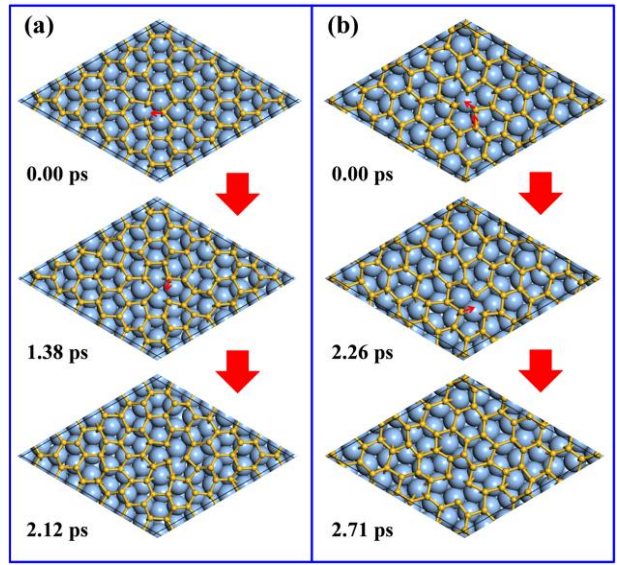

**Figure 6.** Snapshots from the AIMD simulation of silicene monolayers on Ag(111) at 550 K: (a) 4×4 silicene; (b) $\sqrt{13} \times \sqrt{13}$ silicene.

## 2.4 Formation energies and concentrations of defects

The concentration of defects in a material directly determines its fundamental properties and is thus a critical concern for device applications. The probability that a point defect occurs at a given site is proportional to the Boltzmann factor for thermal equilibrium: $P = exp(-E_{form}/k_BT)$, where $E_{form}$ is the formation energy of the point defect, $k_B$ is the Boltzmann constant, $T$ is the temperature.[63] As discussed above, for a given kind of point defect, its formation energy is position dependent. Therefore, for a given kind of point defect, the total number of defects in one unit cell is the sum of the probability $P$ for this kind of point defect occurring at all possible sites. Thus, the concentration $c$ for a given type of defect can be estimated as

$$c = \sum_i \exp\left(-\frac{E_{form}^i}{k_BT}\right)/S \qquad (3),$$

where $E_{form}^i$ is the formation energy of one kind of defect at site $i$ in one unit cell, and $S$ is the area of one unit cell. The formation energies for various point defects in the two silicene superstructures are summarized in Table I. The formation energies for SW defects in $\sqrt{13} \times \sqrt{13}$ silicene (0.815 – 1.264 eV) are lower than those in 4×4 silicene (1.354 – 1.544 eV), indicating a higher concentration of SW defects in $\sqrt{13} \times \sqrt{13}$ silicene. It is noteworthy that the formation energies for Si-SV and Si-DV defects in $\sqrt{13} \times \sqrt{13}$ silicene are extremely low (down to 0.052 eV and 0.079 eV, respectively), suggesting that epitaxial $\sqrt{13} \times \sqrt{13}$ silicene on Ag(111) would be very defective (which will be further discussed later). For both 4×4 and $\sqrt{13} \times \sqrt{13}$ silicene sheets, the formation energy of a Si-DV is much lower than that of two Si-SVs, indicating that two Si-SV defects would coalesce into one Si-DV defect by overcoming a moderate energy barrier (0.47 eV for 4×4 silicene and 0.76 eV for $\sqrt{13} \times \sqrt{13}$ silicene). In the experimental STM image of $\sqrt{13} \times \sqrt{13}$ silicene (Figure 7), there are abundant big black holes and lines of lost bright points, which might originate from coalescing of Si-SV or Si-DV defects. Moreover, the distribution of bright points in the STM image is inhomogeneous due to the easy merging of SVs via diffusion.

**Table I.** Formation energies (in unit of eV) for various point defects in 4×4 and $\sqrt{13} \times \sqrt{13}$ silicene superstructures.

|  | SW | SV | DV | Si-adatom |
|---|---|---|---|---|
| 4×4 | 1.354 – 1.544 | 0.731 – 1.059 | 0.890 – 1.163 | 0.705 – 0.894 |
| $\sqrt{13} \times \sqrt{13}$ | 0.815 – 1.264 | 0.052 – 0.534 | 0.079 – 0.839 | 0.721 – 1.210 |

The equilibrium concentrations of various defects in the two silicene superstructures are estimated by Eq. (3) and listed in Table II. According to Eq. (3), since the concentration depends on the growth temperature of silicene on the Ag(111) surface, which is around 500 K, the temperature in Eq .(3) is set to be 500 K. We first discuss the equilibrium concentration of defects in 4×4 silicene. The equilibrium concentration of SW defect is as small as 14 cm$^{-2}$, which means that there is only one SW defect in a large area of 7 mm$^2$ on average. Thus, the SW defect can hardly be observed in experiments. Compared to SW, the equilibrium concentrations of SV, DV, and Si adatom are larger. The estimated concentration of SV is $4.9 \times 10^7$ cm$^{-2}$, which is much larger than that of DV ($4.6 \times 10^5$ cm$^{-2}$). Si-SV diffuses very fast during the growth process, however, and two SVs can coalesce into one DV to lower the energy. As a result, the final equilibrium concentration of DV should be larger than $4.6 \times 10^5$ cm$^{-2}$, and the concentration of SV should be smaller than $4.9 \times 10^7$ cm$^{-2}$. Nevertheless, SV and DV in 4×4 silicene have relatively small equilibrium concentrations but should still be observed occasionally. The estimated concentration of Si-ad is $2.9 \times 10^7$ cm$^{-2}$, which is comparable to that of SV. Actually, in our experiments we do observe SV, DV, and Siad defects, but not SW defects.

**Table II.** Equilibrium concentrations for various point defects in 4×4 and $\sqrt{13} \times \sqrt{13}$ silicene superstructures in unit of cm$^{-2}$.

|  | SW | SV | DV | Si-adatom |
|---|---|---|---|---|
| 4×4 | 14 | $4.9 \times 10^7$ | $4.6 \times 10^5$ | $2.9 \times 10^7$ |
| $\sqrt{13} \times \sqrt{13}$ | $3.1 \times 10^6$ | $4.4 \times 10^{13}$ | $5.0 \times 10^{13}$ | $2.5 \times 10^7$ |

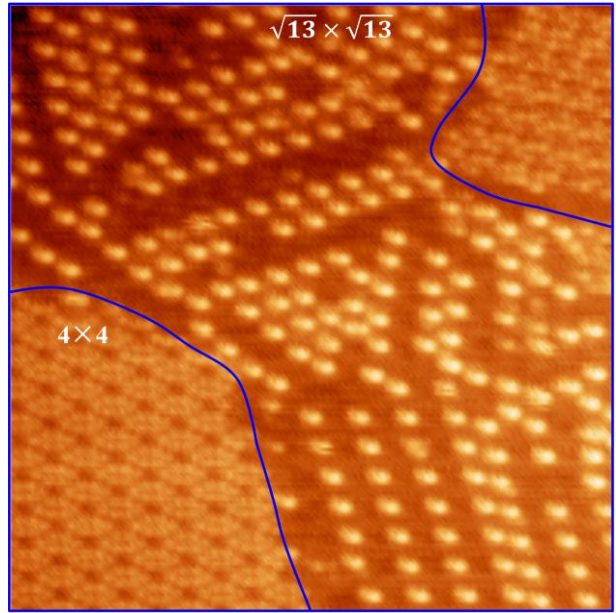

**Figure 7.** Large scale STM image showing 4×4 and $\sqrt{13} \times \sqrt{13}$ silicene from experiment ($V_{bias} = -1.5$ V, $I_{tip} = 4$ nA).

Now, let's turn to $\sqrt{13} \times \sqrt{13}$ silicene. As shown in Table II, the equilibrium concentration of SW defects in $\sqrt{13} \times \sqrt{13}$ silicene ($3.1 \times 10^6$ cm$^{-2}$) is much larger than that in 4×4 silicene. It is still hard to find a SW defect, since only one SW defect exists in an area of 33.3 μm$^2$. Note that SV and DV have very high concentrations of about $5.0 \times 10^{13}$ cm$^{-2}$ in $\sqrt{13} \times \sqrt{13}$ silicene, which means that there should be one SV or DV in every 2 nm$^2$. The large concentration of vacancy defects is clearly reflected in the experimental STM image in Figure 7. Combining the STM observations with the above theoretical results, we conjecture that the observed big black holes and lines of black holes correspond to vacancy clusters from coalescence of Si-SV and Si-DV defects. Consequently, the total concentration of vacancy defects can be obtained by counting the number of missing bright points in the STM image. In such a way, the concentration of missing bright points is $4.1 \times 10^{13}$ cm$^{-2}$ for $\sqrt{13} \times \sqrt{13}$ silicene in Figure 7, which is close to the equilibrium concentration of SV or DV calculated from Eq. (3) in Table II. The accordance between the calculation and experiment demonstrates that our theoretical method is reliable for describing the defect concentration qualitatively. The Si-adatom in $\sqrt{13} \times \sqrt{13}$ silicene has a concentration of $2.5 \times 10^7$ cm$^{-2}$, which is nearly the same as that in 4×4 silicene. Nevertheless, we have not observed Si-adatom in $\sqrt{13} \times \sqrt{13}$ silicene in our experiments as yet.

Based on the discussion above, we can see that the equilibrium concentrations of point defects are generally very small in 4×4 silicene, but they could be rather large in $\sqrt{13} \times \sqrt{13}$ silicene. This nicely explains our experimental observation that 4×4 silicene appears perfect, while $\sqrt{13} \times \sqrt{13}$ silicene looks defective (Figure 7). In addition, in 4×4 silicene, Si-adatom has a concentration comparable to that of Si-SV, and Si-adatom can diffuse very fast throughout the entire silicene sheet, with a diffusion barrier of about 0.4 eV. When a Si adatom diffuses to the site of a Si-SV, the SV can capture the Si adatom and the defect is thus healed. Therefore, appropriate annealing may help improve the quality of epitaxial 4×4 silicene. Generally, 4×4 silicene should be fabricated for future nanoelectronic devices due to its high quality and easy self-healing capability with respect to defects.

## 3. Conclusion

To summarize, the morphologies and energetics of various point defects in epitaxial silicene on Ag(111) surfaces have been systematically investigated using atomistic first-principles calculations combined with experimental scanning tunneling microscopy. The atomic structures for these point defects observed in experimental STM are identified with the aid of DFT calculations. Both cNEB calculations and AIMD simulations demonstrate that Si-SV can diffuse very fast in both 4×4 and $\sqrt{13} \times \sqrt{13}$ silicene sheets at 500 K. Thus, two SVs would coalesce into one DV via diffusion to lower energy. Moreover, Si-adatom also diffuses very fast in 4×4 silicene at 500 K and thus can heal Si-SV defects under appropriate annealing. At 500 K, the equilibrium concentration of SW defects in 4×4 silicene is as low as 14 $cm^{-2}$. The concentrations of SV, DV, and Si-ad defects are also very small, i.e. $4.9 \times 10^7$ $cm^{-2}$, $4.6 \times 10^5$ $cm^{-2}$, and $2.9 \times 10^7$ $cm^{-2}$, respectively. On the contrary, the estimated concentrations of SV and DV defects in $\sqrt{13} \times \sqrt{13}$ silicene are as high as $4.4 \times 10^{13}$ $cm^{-2}$ and $5.0 \times 10^{13}$ $cm^{-2}$. In other words, there would be one single or double vacancy in every 2 $nm^2$ area. The large concentration of point defects and easy diffusion and coalescence of Si-SV nicely explains the defective appearance of $\sqrt{13} \times \sqrt{13}$ silicene in experiments. Therefore, epitaxial 4×4 silicene is thought to be most suitable monolayer silicene phase for future device applications due to the small amount of defects.

## 4. Methods

All first-principles calculations were carried out using the Vienna Ab Initio Simulation Package

(VASP) based on DFT.[64] The electron-ion interactions were described by the projector augmented wave (PAW) potentials.[65] The Perdew-Burke-Ernzerhof (PBE) functional within the generalized gradient approximation (GGA)[66] was adopted. A kinetic energy cutoff of 400 eV for the plane wave basis and a convergence criterion of $10^{-4}$ eV for the total energies were adopted.

The Ag(111) surface was modeled by a three-layer slab model with a vacuum space of more than 12 Å, which was cleaved from bulk face-centered-cubic (fcc) silver with the experimental lattice constant of 2.89 Å. With fixed supercell parameters, the three-layer slab model was further relaxed, with the bottom layer fixed to mimic a semi-infinite solid. Here, we built two selected silicene superstructures on the Ag(111) surface, i.e. 3×3 silicene on a 4×4 Ag(111) surface and $\sqrt{7} \times \sqrt{7}$ silicene on $\sqrt{13} \times \sqrt{13}$ Ag(111) surface, by compressing the silicene lattice slightly to fit the metal surface, following our previous work.[57] To simulate defective silicene supported on Ag substrate, one point defect was created in a 2×2 supercell of a silicene@Ag(111) superstructure to avoid the interactions between adjacent periodic images of the defects in the lateral directions. The lattice constants of our simulation supercells were 23.12 Å and 20.834 Å for 4×4 and $\sqrt{13} \times \sqrt{13}$ silicene, corresponding to $Si_{72}Ag_{192}$ and $Si_{56}Ag_{156}$, respectively. The STM images were simulated by using the Tersoff-Hamann approximation[67] with a constant height of 2 Å above the buckled-up Si atoms. Different bias voltages were tested and found to have virtually no influence on the simulated STM images. Thus, the bias voltage was set to be −1.5 eV for all STM simulations in this paper.

All samples used in this work were fabricated in a preparation chamber supplied with a low-temperature STM/scanning near-field optical microscopy system (LT-STM-SNOM, SNOM1400, Unisoku Co.), as reported elsewhere.[28, 68-69] Clean Ag(111) substrates were prepared by argon ion sputtering and annealed at 800 K for several cycles. The silicene monolayers were then grown on the Ag(111) surfaces by evaporation of silicon from a heated silicon wafer. All the measurements were carried out in ultrahigh vacuum (UHV) at 77 K. Pt/Ir tips were calibrated on a silver surface before STM measurements.

**Acknowledgements**

This work was supported by the National Natural Science Foundation of China (11134005, 11375228, 11574040, 11575227), Project 1G2009312311750101 of the Chinese Academy of

Sciences, the Australian Research Council (ARC) through a Discovery Project (DP140102581), the University of Wollongong through a University Research Council (URC) Small Grant in 2014, and ARC Linkage Infrastructure, Equipment and Facilities (LIEF) grants (LE100100081 and LE110100099).

# Supporting Information

# Points defects in epitaxial silicene on Ag(111) surface

Hongsheng Liu[1], Haifeng Feng[2], Yi Du[2*], Jian Chen[3], Kehui Wu[3], Jijun Zhao[1,4*]

[1] *Key Laboratory of Materials Modification by Laser, Ion and Electron Beams (Dalian University of Technology), Ministry of Education, Dalian 116024, China*

[2] *Institute for Superconducting and Electronic Materials (ISEM), University of Wollongong, Wollongong, New South Wales 2525, Australia*

[3] *Beijing National Laboratory for Condensed Matter Physics and Institute of Physics, Chinese Academy of Sciences, Beijing 100190, China*

[4] *Beijing Computational Science Research Center, Beijing 100084, China*

---

[*] Corresponding authors. Email: zhaojj@dlut.edu.cn (J. Zhao), ydu@uow.edu.au (Y. Du)

**S1. Structures, formation energies and STM images of various defective 4×4 silicene superstructures.**

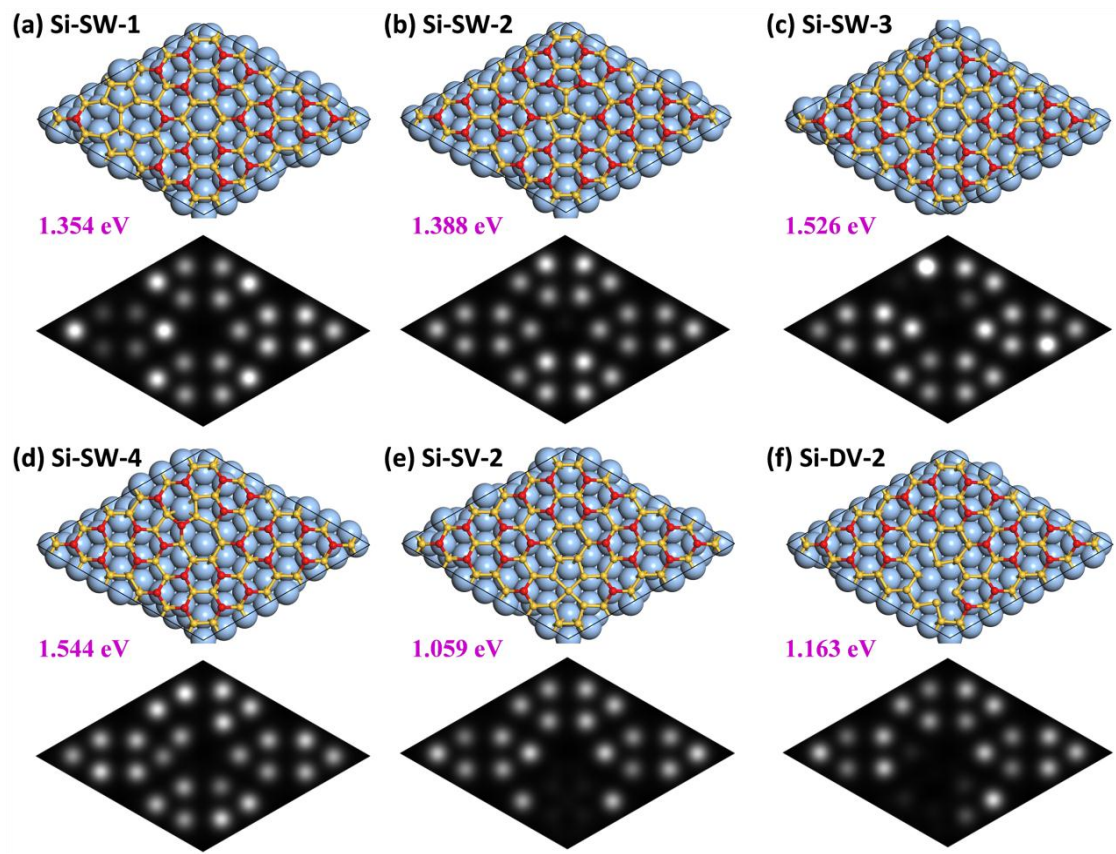

**Figure S1.** Atomic structures, formation energies and simulated STM images of various defective 4×4 silicene superstructures. The sky-blue, yellow and red balls represent Ag atoms, buckled down Si atoms and buckled up Si atoms respectively. The black rhombus indicates the supercell we used in our calculations. Formation energies are labeled below each structure.

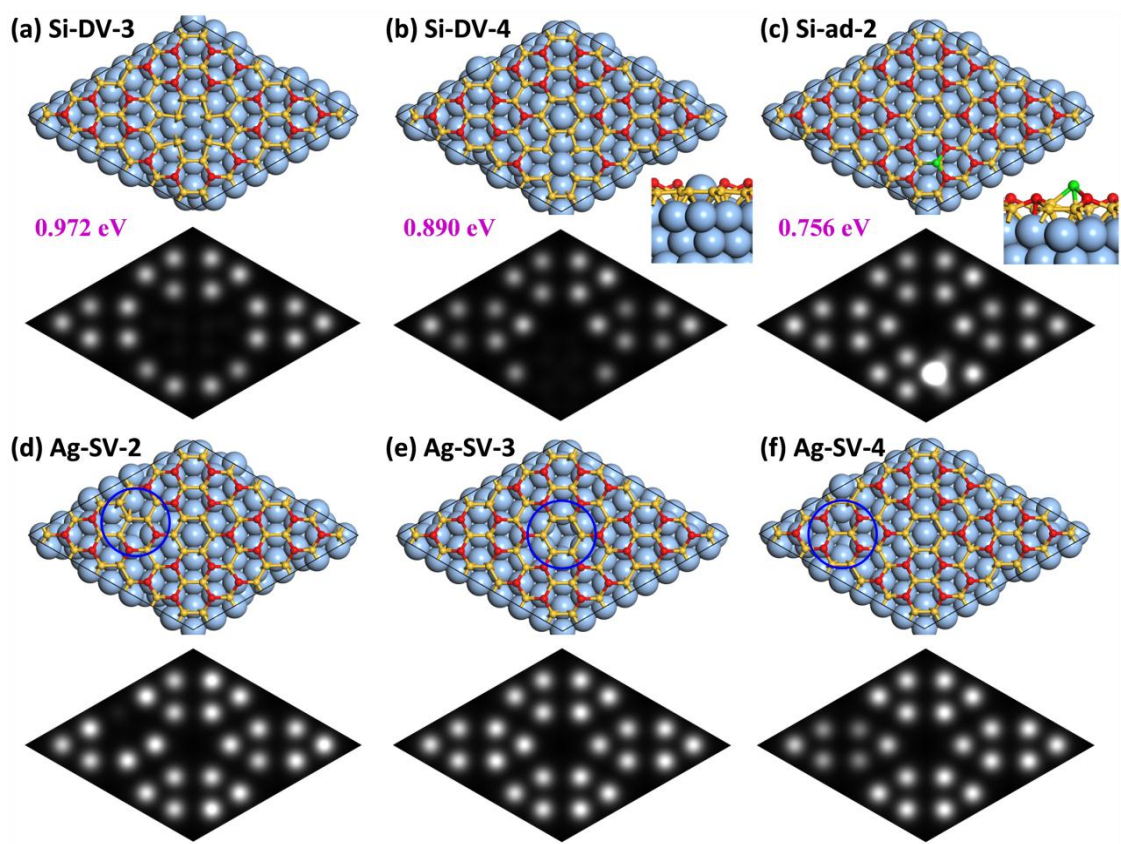

**Figure S2.** Atomic structures, formation energies and simulated STM images of various defective 4×4 silicene superstructures. The small picture in (c) is the local side view around the Si adatom. The sky-blue, yellow, red and green balls represent Ag atoms, buckled down Si atoms, buckled up Si atoms and Si adatoms respectively. The dark blue circle indicates the position of the Ag vacancy. The black rhombus indicates the supercell we used in our calculations. Formation energies are labeled below each structure.

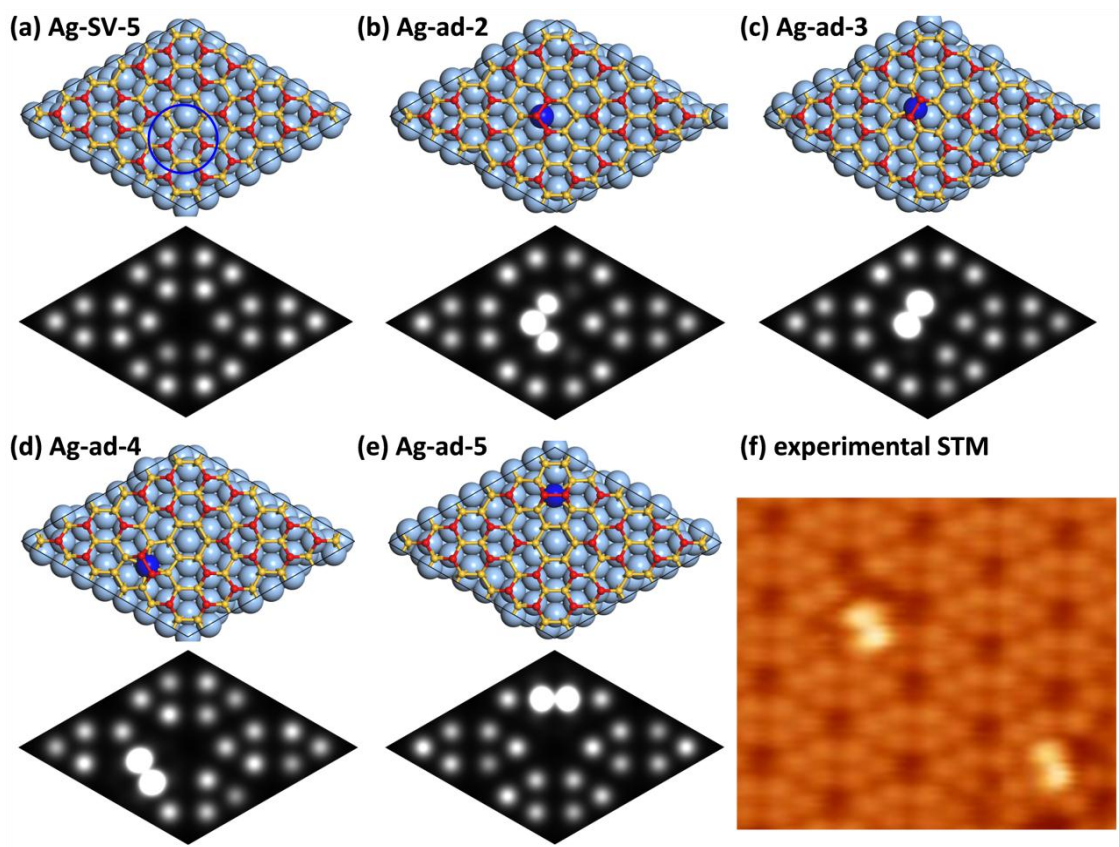

**Figure S3.** Atomic structures and simulated STM images of various defective 4×4 silicene superstructures as well as experimental STM image. The sky-blue, yellow and red balls represent Ag atoms, buckled down Si atoms and buckled up Si atoms respectively. The dark blue circle indicates the position of the Ag vacancy. The black rhombus indicates the supercell we used in our calculations. The bias voltage for experimental STM images is 1.2 eV.

Si-SW-1, which is the most stable configuration for SW defect in 4×4 silicene, has formation energy of 1.354 eV. The bond rotation between one buckled up Si atom and one buckled down Si atom (Si-SW-3 and Si-SW-4) would introduce larger distortion in silicene sheet than that between two buckled down Si atoms (Si-SW-1 and Si-SW-2) and thus results in larger formation energy. Si-SW defects result in lack of bright point (Si-SW-1 and Si-SW-3) or distortion of bright point pattern (Si-SW-4) in the STM image. Si-SW-2 has nearly no influence on the STM image. Si-DV defect in different positions results in similar structure except for detailed buckled pattern and formation energy, as shown in Figure S1 and S2. Note that for Si-DV-4, which is the most stable configuration for Si-DV, one Ag atom below the DV defect in the substrate is lift up. However, the lift up Ag atom cannot be seen in the STM image due to its lack of LDOS around the Fermi level. Si-ad-2 (Figure S2c), which possesses a dumbbell configuration similar to that in free-standing silicene[1], has a formation energy about 0.05 eV higher than that for Si-ad-1.

For Ag-SV-2 shown in Figure S2d, one original buckled up Si atom goes down due to the Ag-SV, which is the same to Ag-SV-1. When the missing Ag atom is right underneath the hollow site of the silicene sheet (Ag-SV-3 in Figure S2e) or underneath the buckled down Si atom (Ag-SV-4, Ag-SV-5 in Figure S2f and Figure S3a), the buckling pattern of 4×4 silicene remains unchanged. The local buckling pattern of 4×4 silicene will also be severely influenced by an Ag adatom on the Ag(111) surface. When the Ag adatom is right beneath one Si atom, the Ag adatom will lift up the Si atoms above and introduce three big bright points in STM image, as shown in Figure S3b. When the Ag adatom is under a Si-Si bond (Ag-ad-3, Ag-ad-4, and Ag-ad-5), it would induce a dumbbell composed of two big bright points in the STM image, as shown in Figure S3c-e. Moreover, dumbbells have been found experimentally (Figure S3f), although no exact match could be found between the simulated and experimental STM images.

**S2. Structures, formation energies and STM images of various defective $\sqrt{13}\times\sqrt{13}$ silicene superstructures.**

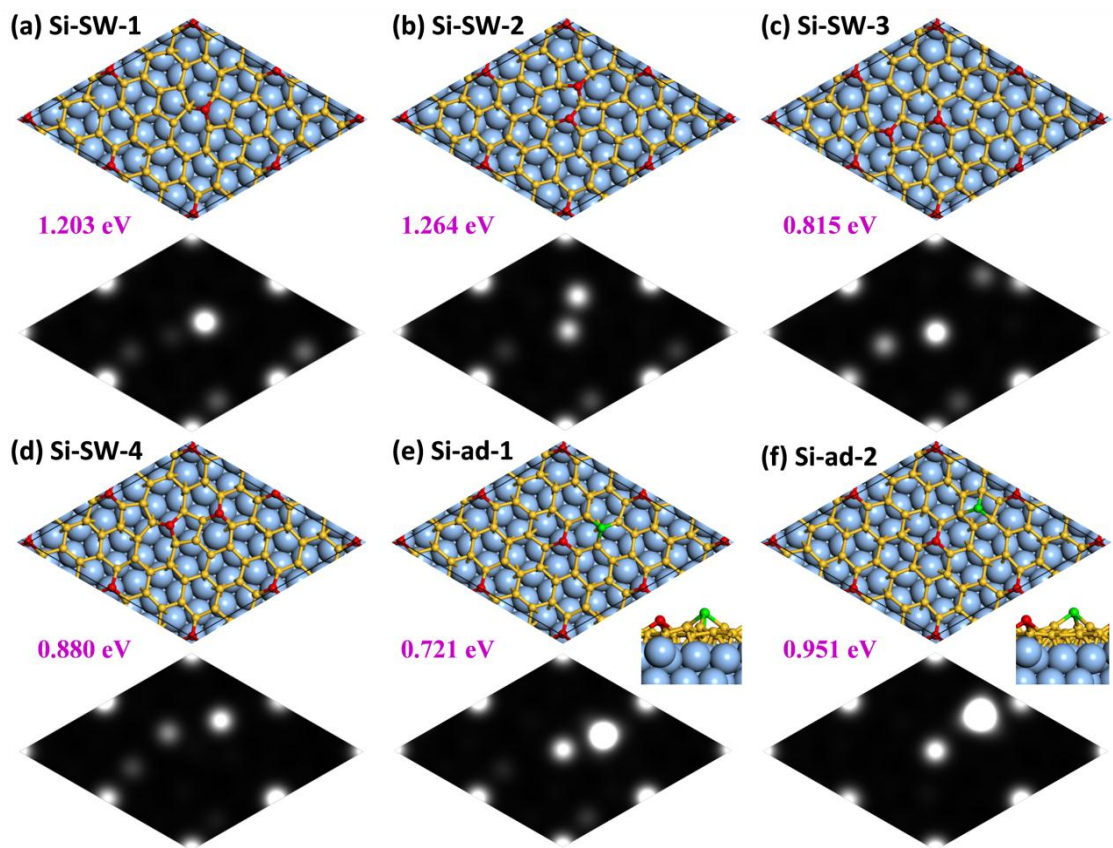

**Figure S4.** Atomic structures and simulated STM images of various defective $\sqrt{13}\times\sqrt{13}$ silicene superstructures. The sky-blue, yellow, red and green balls represent Ag atoms, buckled down Si atoms, buckled up Si atoms and Si adatoms respectively. The black rhombus indicates the supercell we used in our calculations.

For Si-SW defects in $\sqrt{13}\times\sqrt{13}$ silicene, different positions of bond rotation result in different local buckling patterns and thus distinct STM images (Figure S4a-d). For Si-SW-1, one bright point in the middle deviate a little from its original position due to the bond rotation. Si-SW-2 and Si-SW-3 introduce more bright points in the STM images. For Si-SW-4, the original bright point in the middle disappears and two bright points arise aside instead. Si-ad-1 (Figure S4e), which possesses a dumbbell configuration similar to that in freestanding silicene[1], is the most stable configuration for Si adatom with formation energy 0.23 eV lower than that of Si-ad-2 (Figure S4f). Their simulated STM images, although not experimentally recognized yet, resemble each other to some extent (Figure S4e, f).

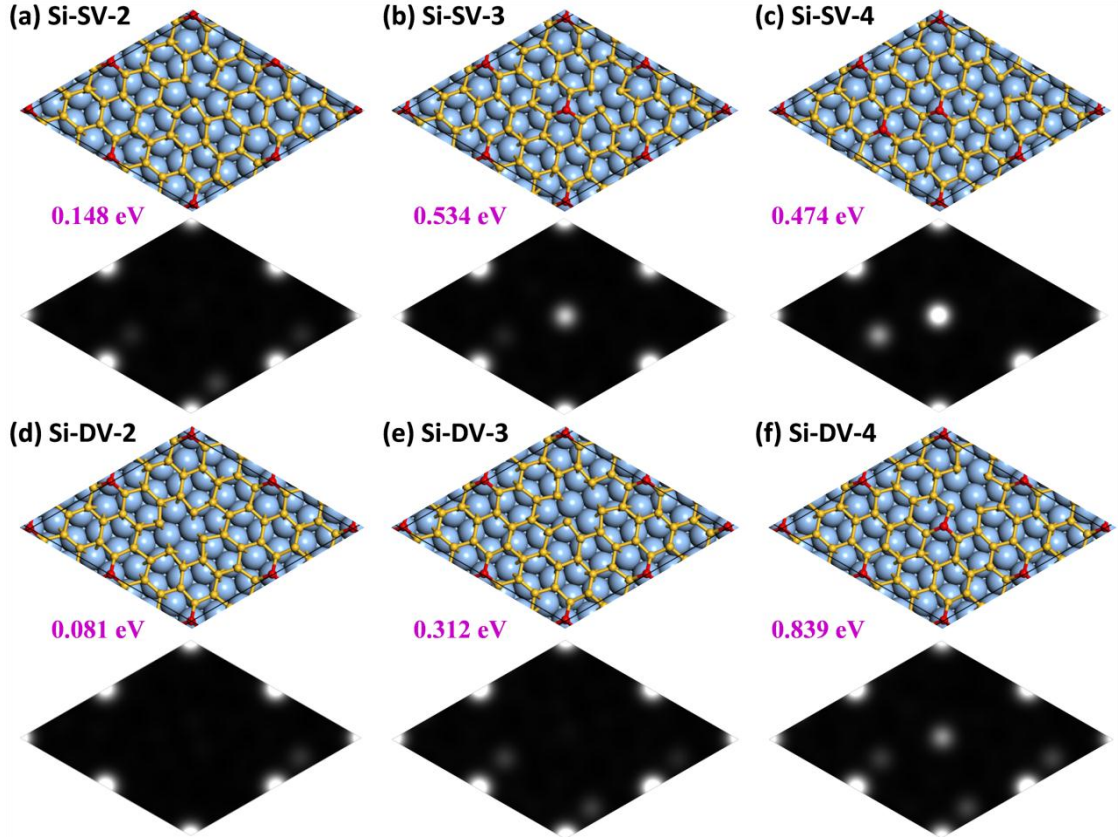

**Figure S5.** Atomic structures and simulated STM images of various defective $\sqrt{13}\times\sqrt{13}$ silicene superstructures. The sky-blue, yellow and red balls represent Ag atoms, buckled down Si atoms and buckled up Si atoms respectively. The black rhombus indicates the supercell we used in our calculations.

Si-SV-2 results in one missing bright point in the STM image which is the same to Si-SV-1. In contrast, Si-SV-3 has nearly no influence on the STM image since the buckle pattern of silicene remains unchanged after the detachment of one buckled-down silicon atom. The STM image of Si-SV-4 is very different from that of perfect silicene, one original bright point missing and one new bright point appearing. Si-DV-2 and Si-DV-3 give the same STM image that one bright point is missing. Si-DV-4, however, has little influence on the STM due to little influence on the buckle pattern of silicene. The formation energies for Si-SV-3, Si-SV-4 and Si-DV-4 are much higher those for Si-SV-1 and Si-DV-1 since the missing atoms are buckled down atoms.

## S3. Diffusion paths of Si-SV and Si-ad in $\sqrt{13} \times \sqrt{13}$ silicene.

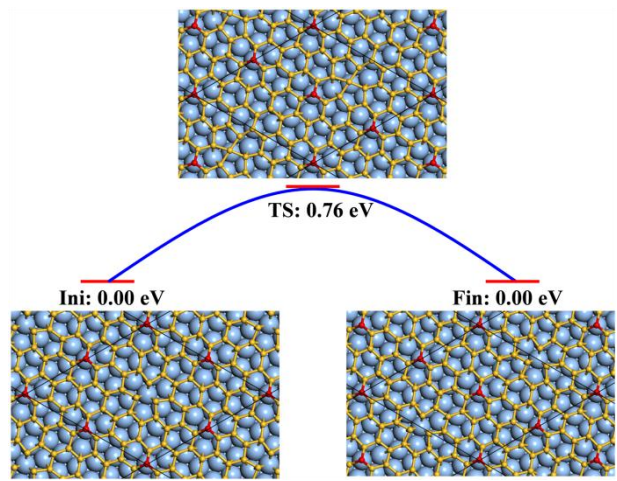

**Figure S6.** Schematic plots for diffusion of a SV defect in $\sqrt{13} \times \sqrt{13}$ silicene. The sky-blue, yellow and red balls represent Ag atoms, buckled down Si atoms and buckled up Si atoms respectively. The black rhombus indicates the supercell we used in our calculations.

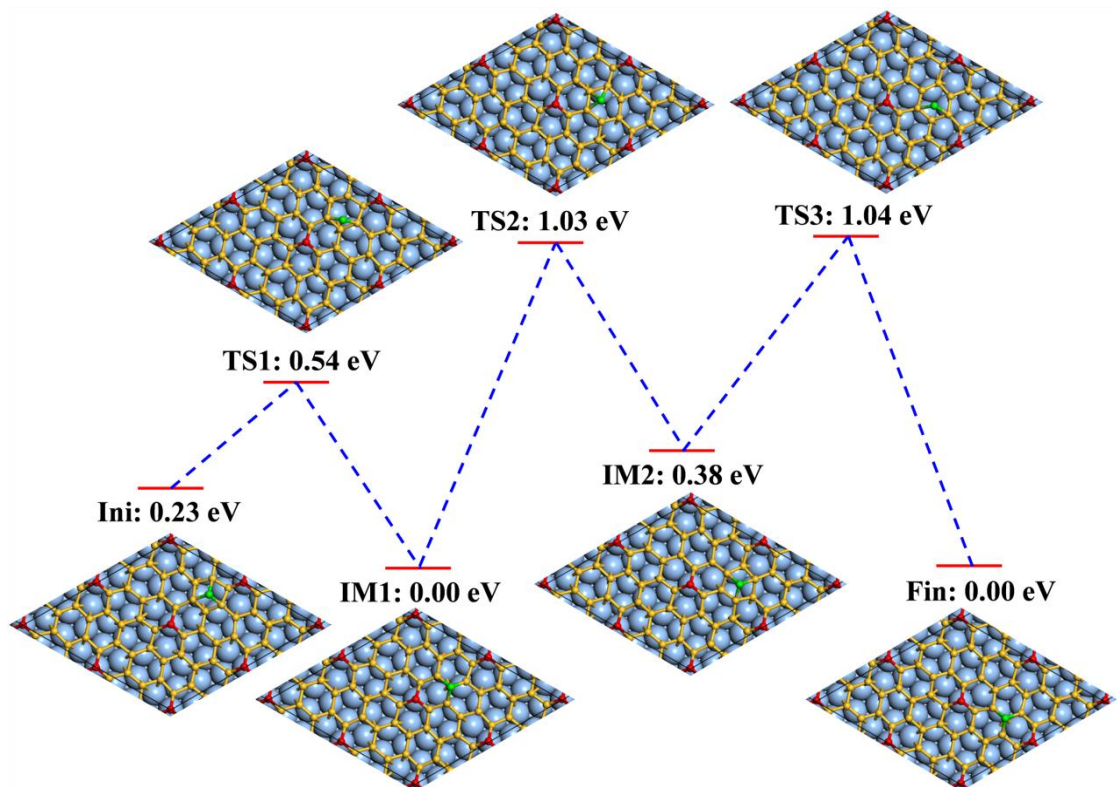

**Figure S7.** Schematic plots for diffusion of one Si adatom in $\sqrt{13} \times \sqrt{13}$ silicene. The sky-blue, yellow, red and green balls represent Ag atoms, buckled down Si atoms, buckled up Si atoms and Si adatoms, respectively. The black rhombus indicates the supercell we used in our calculations.

**References**

[1] J. Gao, J. Zhang, H. Liu, Q. Zhang, J. Zhao, *Nanoscale* **2013,** *5*, 9785-9792.